# Energy Efficiency: The New Holy Grail of Data Management Systems Research

Stavros Harizopoulos
HP Labs
Palo Alto, CA
stavros@hp.com

Mehul A. Shah
HP Labs
Palo Alto, CA
mehul.shah@hp.com

Justin Meza
UCLA
Los Angeles, CA
justin.meza@ucla.edu

Parthasarathy Ranganathan
HP Labs
Palo Alto, CA
partha.ranganathan@hp.com

## ABSTRACT

Energy costs are quickly rising in large-scale data centers and are soon projected to overtake the cost of hardware. As a result, data center operators have recently started turning into using more energy-friendly hardware. Despite the growing body of research in power management techniques, there has been little work to date on energy efficiency from a data management software perspective.

In this paper, we argue that hardware-only approaches are only part of the solution, and that data management software will be key in optimizing for energy efficiency. We discuss the problems arising from growing energy use in data centers and the trends that point to an increasing set of opportunities for software-level optimizations. Using two simple experiments, we illustrate the potential of such optimizations, and, motivated by these examples, we discuss general approaches for reducing energy waste. Lastly, we point out existing places within database systems that are promising for energy-efficiency optimizations and urge the data management systems community to shift focus from performance-oriented research to energy-efficient computing.

## Categories and Subject Descriptors

H.2.4 [**Database Management**]: Systems — *relational databases; query processing.* H.3.4 [**Information Storage and Retrieval**]: Systems and Software — *performance evaluation (efficiency and effectiveness).*

## General Terms

Algorithms, Performance, Design.

## Keywords

Data management systems, database systems, energy efficiency, power management.

## 1. INTRODUCTION

Growing demands for information processing have lead to a demand for cheaper, faster, and larger data management systems. At the same time, an important and growing component of the total cost of ownership for these systems is power and cooling [Bar05]. A recent report by the Environmental Protection Agency shows that data center power consumption in the US doubled between 2000 and 2006, and will double again in the next five years [EPA07]. Uncontrolled energy use in data centers also has negative implications on density, scalability, reliability, and the environment. These trends are raising awareness across multiple disciplines to optimize for energy use in data centers.

Recent work in computer architecture and power management has called for system designs and provisioning of hardware components that take power consumption under consideration. The goal in these approaches is to design systems that will be highly energy efficient at the peak performance point (maximum utilization) and remain energy efficient as load fluctuates or drops. The notion of *energy proportionality* [BH07] characterizes exactly that: servers should use no power when not used and power only in proportion to delivered performance or system utilization. Thus, servers would offer constant *energy efficiency*, i.e., the ratio of performance to power, at all performance levels.

In this paper, we argue that energy-friendly and energy-proportional hardware is only part of the answer. Software choices can be particularly effective, especially in database management systems that permit broad choices due to physical data independence and query optimization. We expect the opportunity for software-level optimizations to be even larger with the recent trend of increased heterogeneity at all levels, from clusters to storage hierarchies and CPUs. We demonstrate the efficacy of software choices in database systems through two simple examples. In the first example, we experiment with a system which was configured similarly to an audited TPC-H server and show that making the right decision in physical design can improve energy efficiency. The second example uses a relational scan operator as a basis to demonstrate that optimizing for performance is different from optimizing for energy efficiency.

Motivated by these examples, we discuss potential approaches for reducing energy waste. We identify three such general approaches, each with an increased level of complexity: (a) adjust existing system-wide configuration knobs and query optimization parameters to account for energy costs, (b) consolidate resource use in both time and space to facilitate powering down unused hardware components, and (c) revisit and redesign data structures, algorithms, and policies, from a whole-system perspective, to improve energy efficiency. With these approaches in mind, we examine several areas of database systems that are ripe for energy efficiency optimizations. These areas range from physical design to storage and buffer management, and from query optimization and query processing to new architectures and special purpose engines.

Our aim in this paper is to bring awareness to the database systems community about the important opportunities for research on energy-efficient data management software. We start our discussion of the problem by presenting trends and implications of rising energy use (Section 2), and motivate the potential that database software has for affecting energy efficiency. Section 3 contains our two examples that further highlight the potential for software-based energy optimizations. In Section 4, we describe three general approaches for reducing energy waste, and in Section 5, we discuss promising areas in data management systems for improving energy efficiency. We conclude in Section 6.

## 2. THE NEED FOR ENERGY EFFICIENCY

In this section, we first define and motivate the need for energy efficiency in data management systems (Sections 2.1 and 2.2), and then highlight current findings and trends that will shape the scope of future solutions (Sections 2.3 and 2.4).

### 2.1 What is Energy Efficiency

Energy is the physical currency used for accomplishing a particular task, e.g., moving a car, lighting a room, or even performing a computation. It can take many forms such as electrical, light, mechanical, nuclear, and so on, and can be converted from one form to another. For computing systems and data centers, energy is delivered as electricity. Power is the instantaneous rate of energy use, or equivalently, energy used for a task is the product of average power used and the time taken for the task:

$$Energy = AvgPower \times Time$$

The typical units for energy and power are Joules and Watts, respectively, and 1 Joule = 1 Watt x 1 second.

For computer systems, we roughly define energy efficiency as the ratio of "computing work" done per unit energy. It is analogous to the miles per gallon metric for automobiles. This metric varies from application to application since the notion of work done varies. For example, it might be transactions/Joule for OLTP systems, and searches/Joule for a search engine. Energy efficiency is also equivalent to the ratio of performance, measured as the rate of work done, to power used:

$$EE = \frac{WorkDone}{Energy} = \frac{WorkDone}{Power \times Time} = \frac{Perf}{Power}$$

For fixed amount of work, maximizing energy efficiency is the same as minimizing energy. Thus, unlike performance optimization, we can improve energy efficiency by reducing power, time, or both.

### 2.2 Energy Cost vs. Data Management Needs

Energy use in data centers is a growing problem and a key concern for data center operators and IT executives. A recent study by the Environmental Protection Agency (EPA) shows that 60 billion kWh, or 1.5% of the total US energy use in 2006, was used to power data centers, and this use is expected to nearly double by 2010 [EPA07]. Koomey also observes a similar trend and estimates $2.7 billion was spent in the US and $7.2 billion was spent worldwide to power and cool servers in 2005 [Koo07]. Moreover, studies show that every 1W used to power servers requires an additional 0.5W to 1W of power for cooling equipment [PBS+03]. Although energy costs and breakdowns vary depending upon the installation, analysts predict that energy costs will eventually outstrip the cost of hardware [Bar05]. Data center operators will, therefore, need to adjust their pricing structure accordingly to reflect these rising energy costs.

Besides the cost of electricity, energy use by computing equipment has implications on data center density, reliability, and the environment. Racks in data centers are provisioned to deliver a certain capacity in order to properly power and cool the servers. As power consumed by servers increases, many racks end up going empty. Even when racks deliver enough power, often cooling is the limitation, since undercooled equipment exhibits higher failure rates. Thus, energy-use limits achievable data-center level scalability. Finally, energy use in data centers is starting to prompt environmental concerns of pollution and excessive load placed on local utilities [PR06]. As a result of these trends, governmental agencies (e.g., EPA, US Congress, Intelligent Energy Europe, TopRunner) are actively seeking to regulate enterprise power. Recently, a new industrial consortium, GreenGrid, has been formed to address energy efficiency in data centers.

At the same time, the demand for cheaper, faster, and larger data management systems continues to grow. Some systems in data centers support traditional data management tasks like transaction processing and business intelligence. The business intelligence market is a multi-billion dollar market and still seeing double-digit growth rates [VMW+08]. With the growth of the internet sector, data centers are also running non-traditional data management workloads such as search, multimedia storage and delivery, web analytics, and more recently, cloud computing. To accommodate their rapidly growing datasets and workload, internet-sector companies like Google are increasingly concerned with energy use [Bar05]. For example, such companies are starting to build data centers close to electric plants in cold-weather climates [MH06]. Ultimately, however, to tackle the energy problem in data centers while continuing to fuel the demand for data management resources, we will need to shift our focus from optimizing data management systems for pure performance to optimizing for energy efficiency.

### 2.3 Initial Reactions to Energy Concerns

To tackle the energy-related concerns, there has been a plethora of work from the architecture and power management communities. The previous work spans all levels from chips to data centers [Ran09]. At the chip level, designers have considered techniques such as dynamic voltage and frequency scaling (DVFS), clock routing optimizations, low-power logic, asymmetric multi-cores,

CIDR Perspectives 2009

and so on. At the platform level, there has been work on reducing power supply inefficiencies as well as power optimizations for the memory hierarchy. For example, researchers have suggested strategies for dynamically turning off DRAM, disk speed control, and disk spin down [ZCT+05]. More recently, there has been an emphasis on cluster-level optimizations: shifting workloads and power budgets by considering power and temperature constraints across multiple domains and the data center as a whole. Finally, to complement these compute-based techniques, there have been improvements in data center cooling. Unfortunately, most of this past work has been application and database agnostic.

Recent work on energy-efficiency optimizations for data management workloads has proposed the JouleSort [RSR+07] and SPEC-Power benchmarks. These two measure energy efficiency of entire systems that perform data management tasks. Their workloads, however, are quite specialized.

As database system designers, we know that technology inflection points in the past such as virtual memory, deeping of the cache hierarchy, fast networks for clustering, and so on have necessitated a reconsideration in the database domain. Application-agnostic techniques for scaling have hardly helped if not obstructed the performance of database systems.

## 2.4 The Role of Data Management Software

To date, there has been little work on energy efficiency from a data management software perspective. One source of the problem lies with the limited dynamic power range and limited power knobs that most hardware offers today. Analysis of TPC-C [PN08], SPEC-Power and external sort [Riv08] systems, show that most servers offer little power variance from no load to peak use. Thus, for a particular hardware configuration, performance optimizations are the only way to effect energy improvements. Barroso and Holze noticed the same inelasticity in power for Google's servers [BH07], and also found that the CPUs on these servers were mostly between 10-50% utilized. They, therefore, argued for *energy-proportional* systems: servers that use no power when not used and automatically consume power only in proportion to delivered performance, or in their case, system utilization. Such ideal energy-proportional systems would offer constant energy efficiency at all performance levels rather than the best energy efficiency only at peak performance.

Unfortunately, systems or even their components (CPUs, memory, disks, etc.) are hardly energy proportional. Existing components offer only limited controls for trading power for performance. CPUs, for example, offer voltage and frequency scaling, which is a good first step but far from ideal. Memory and disks, whose power contribution can dwarf other components in database systems [PN08], offer almost no power control except for sleep states. They are either on (and at full performance and power) or off, and the transitions can be expensive.

These limited choices in power-performance states prevent data management software from varying power consumption and thereby improving energy efficiency. However, we expect the opportunities for software-level optimization to increase for two reasons. First, there is a push to design components that offer more control over power-performance tradeoffs. For example, we believe it is reasonable to expect that a software module will be able to control which CPU cores in a multicore chip are active at any time.

Second, there is increasing hardware heterogeneity at all levels. At the cluster level, data centers already contain a heterogeneous collection of servers that offer different power-performance tradeoffs because of the technology refresh cycle. Recent work has considered using virtual machine migration and turning off servers to effect energy-proportionality [TWM+08]. Although promising, this approach ignores the disk subsystem. At the platform level, others have suggested blade system designs with an increasing number of choices, e.g., a remote memory blade, compute blades with components from embedded systems, flash blades, and so on [LRC+08]. Finally, at the chip level, processor manufacturers are starting to release heterogeneous CPUs such as IBM's Cell, AMD's Fusion (CPU/GPGPU), and Intel's Larrabee (GPGPU). These trends suggest that software running in data centers will have many more hardware choices with different power-performance characteristics.

Database systems in particular are well equipped to harness these choices due to physical data independence and query optimization. Databases can dynamically choose the hardware based on the workload or consolidate data and processes to effect component-level energy proportionality. Moreover, they can improve energy efficiency even further by choosing among a variety of query processing algorithms that compute the same result but exercise hardware components differently. We explore these and other options for energy-use improvements in the upcoming sections.

## 3. EXAMPLE OPPORTUNITIES

In this section, we describe two examples that show that choices made by database systems can improve energy efficiency independently of improving performance. The first is an experiment that varies the number of disks to affect power consumption for a decision support workload. Using this coarse knob, the system shows a point of diminishing returns for energy efficiency; the most efficient point is not the best performing. The second example shows that algorithms designed for energy efficiency do not necessarily result in the best performance. Although these examples are simple, they illustrate the tradeoffs that we need to consider when optimizing data management systems for energy efficiency.

## 3.1 Example 1: Diminishing Returns

Most database systems running complex workloads can be configured and tuned to show a maximum energy-efficiency point at less than peak performance. The intuition behind this is simple. When configuring a system for high performance, additional instances of any one system component, after a certain point, provide decreasing incremental performance benefit, but add constant amounts of power into the power budget. For example, the 7th disk provides less incremental performance benefit than the 6th disk, more incremental benefit than the 8th disk, and each additional disk contributes the same power. Therefore, in configuring and tuning a system for energy efficiency, one ought to bal-ance system components such that the incremental benefits among all types outweigh the additional power cost.

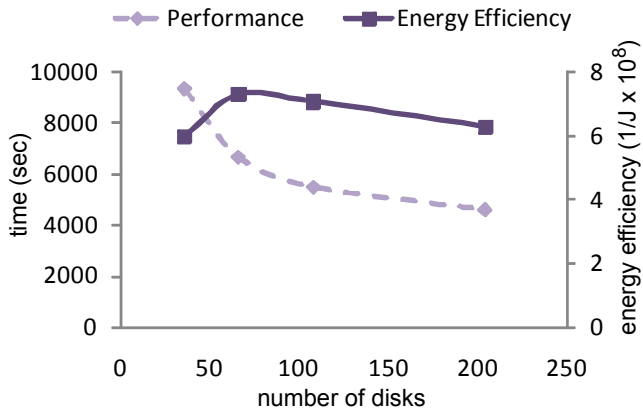

**Figure 1. Time and energy-efficiency vs. number of disks for TPC-H Throughput Test. The points correspond to 36, 66, 108, and 204 disks.**

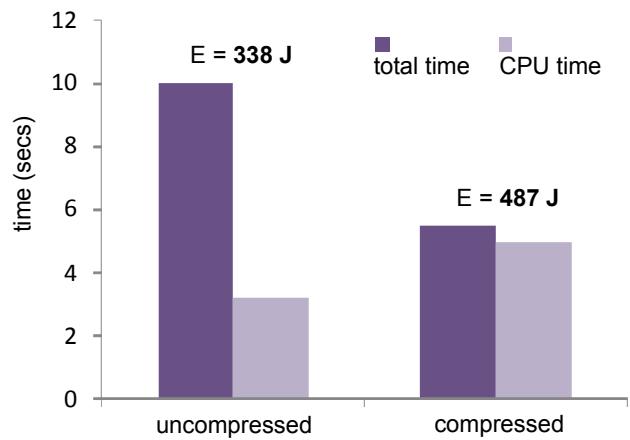

**Figure 2. Scan experiment using one CPU and three flash disks. Compressed data result in a faster query by trading CPU cycles for disk bandwidth, but overall energy consumption increases.**

We demonstrate this effect by exploring the energy-efficiency profile of a high-performance database server running a decision support workload (TPC-H). We found that the most effective means of varying power use in our system was by repartitioning our database across fewer disks, since the disk subsystem consumed more than 50% of the total system power.

The experimental setup is as follows. Although our results are not audited, our system was configured similar to one running an audited TPC-H benchmark result at 300GB scale factor [HP08]. The system consisted of an HP ProLiant DL785 server tray with 8 Quad-Core AMD Opteron processors, 64GB of memory, and between 36-204 SCSI drives (15K RPM, 73GB) connected by SAS to between 2-13 HP StorageWorks MSA70 disk trays. We ran a commercial database system also configured similar to the audited system [HP08] and ran on Microsoft Windows Enterprise Server 2008. We striped the database across all disks in a RAID 5 configuration and used compression to bring the database footprint to about 256GB. Processors were run in the C0 state, and, when idle, the system simply ran the system idle process.

Figure 1 shows the time (performance) and energy efficiency of the system as we varied the number of disks for the "throughput test." The throughput test issues a mixture of TPC-H queries simultaneously from multiple clients to the system. We see that 66 disks was the point of diminishing returns. The percentage gain in performance did not outweigh the percentage gain in power, so energy efficiency started to drop. For this system, the most efficient point offers a 14% increase in efficiency for a 45% drop in performance. The point of diminishing returns and efficiency-performance tradeoff will vary with the workload. Some workloads may be able to use additional resources while others will underutilize them and therefore waste power.

This experiment shows that even with a coarse power control, there is opportunity for improvement. Clearly, the overall benefits also depend on the costs associated with creating or maintaining different partitionings of the database. However, as the number of knobs available and configuration choices that affect power increase, the larger the role the database software will play in optimizing the system for energy efficiency.

### 3.2 Example 2: Algorithm Design

In this example, we show that given energy-proportional components, designing algorithms for energy efficiency is not the same as designing algorithms for performance. We show this by considering a scan with and without compression over a database on flash drives.

Consider a table stored on disk and a simple selection query which scans the entire table and applies a predicate. We draw our numbers for this example from an implementation of a high-performance column-oriented relational scanner from our previous work [HLA+06]. In particular, we pick a query that projects five out of seven attributes of table ORDERS from TPC-H and examine two different configurations, one where the base table is non-compressed and one where it is compressed. The hardware configuration includes one CPU and three SSD flash disks, which are an order of magnitude more energy efficient than regular hard drives.

We reproduce the CPU and total times in Figure 2. When uncompressed, the query is disk-bound: it takes 10 secs to read all data and 3.2 secs for the CPU to process it. By overlapping disk with CPU time, the total time is 10 secs. When compressed, the query becomes mostly CPU-bound, and takes 5.5 secs to run (out of which, 5.1s are CPU time). Clearly, for that particular configuration, if one were to run this query by itself, they would prefer the compressed version of the table, and would observe a speedup of 2x.

With energy efficiency in mind, however, it turns out that the uncompressed table results into a more energy-efficient (but slower) query. The CPU has a power consumption of 90 Watts, while the flash disks together consume only 5 Watts. Therefore, for the uncompressed table, the total energy consumed is (90W x 3.2s + 5W x 10s) = 338 Joules, whereas for the compressed one is (90W x 5.1s + 5W x 5.5s) 487 Joules (assuming that an idle CPU does not consume any power, or, simply, assuming that some other concurrent task is taking up the rest of the CPU cycles).

This simple example points out just how counter-intuitive optimizing for energy efficiency can be. From a performance point of view, the compressed table makes sense: it results in trading 1.9 secs of CPU time for 4.5 secs of disk time, and also overlaps CPU and disk use. However, from an energy efficiency point of view, the compressed table uses a lot more of the CPU. The CPU power (90W) is much larger relative to the power of the Flash drives (5W). Thus, using the additional power to increase performance is not worth it from an energy perspective. This example shows that both power and performance considerations must be evaluated when designing algorithms for energy efficiency.

## 4. APPROACHES FOR REDUCING WASTE

The previous section provided a brief motivation of how optimizing for energy efficiency can differ significantly from optimizing for performance. The space of possible techniques or optimizations for reducing energy waste is obviously larger than what those two examples suggest. In fact, we expect that most approaches that have been applied to date to improve performance, will have a counterpart in the approaches for optimizing for energy efficiency. We highlight this duality between performance and energy optimizations throughout the rest of this section.

The first step to deploying an energy-efficient data management platform is to choose hardware components with good performance-per-watt characteristics. This is analogous to choosing high-performance components (server-grade CPUs, high-RPM disks etc.) when absolute performance is desired. With that as a starting point, we discuss three categories of software-based approaches for reducing wasted energy in data management systems, presented in order of increased complexity. A more general list of power optimizations can be found in [Ran09].

### 4.1 Energy-aware Optimizations

*Approach: Use existing system-wide knobs and internal query optimization parameters to achieve the most energy-efficient configuration for the underlying hardware.*

All modern commercial database systems offer a multitude of knobs, from collecting statistics for query optimization to configuring input parameters to physical designers, and from selecting the degree of parallelization to assigning memory to operators or temporary space. The same way many of those knobs have been tuned to date to increase performance, we expect DBAs to use them to improve energy efficiency. The examples in the previous section demonstrate the necessity in understanding and modelling power cost, so that database administrators and developers can incorporate or apply the model into existing tools.

Choosing the right algorithms and settings for improving energy efficiency entails different tradeoffs than when optimizing for performance, as shown in Section 3.2. Compression techniques, for example, trade off CPU cycles for reduced bandwidth requirements (both disk-to-memory and memory-to-CPU). By turning the focus on energy efficiency, tradeoffs like this one will need to be re-examined.

We expect that query optimization decisions also have the potential to significantly affect energy consumption by producing significantly different query plans. As an example, consider the hash-join operator which has been known to outperform nested-loop join in many occasions, but it relies on using a large chunk of memory for building and maintaining the hash table. From a power perspective, these are "expensive" operations and may tip the balance in favor of nested-loop join in more occasions than before. To improve energy efficiency, query optimizers will need power models to estimate energy costs. There has been a lot of work on modeling power, but simple models may suffice in the same way simple models for device access times work well in practice.

In addition to configuring and tuning a system for maximum energy efficiency in a given hardware configuration, the increased heterogeneity in hardware resources in large data centers will force knob settings and query optimization decisions to be made dynamically, depending on runtime conditions.

### 4.2 Resource Use Consolidation

*Approach: Shift computations and relocate data to consolidate resource use both in time and space, to facilitate powering down individual hardware components.*

Whenever system resources are not used or are partially used, there is an opportunity for saving energy by either allowing other concurrent tasks to utilize the otherwise idle resource or by allowing the resource to enter a suspended or reduced power mode. Ideally, this should be automatically handled by the underlying hardware and/or the operating system. For example, if the disk subsystem is periodically accessed, it should automatically enter into a sleep mode during periods with no activity. Such functionality is necessary for achieving energy-proportional components. However, as we pointed out in Section 2, current-technology components have limited power states and, furthermore, the switching costs across states can easily exceed energy savings.

While we expect hardware components to improve over time in their ability to consume power in proportion to their usage, we believe that software choices and practices will ultimately improve energy-proportionality at the component level. Hardware components will require a certain minimum-length idle period to enter in a suspended mode, and the longer that period is the easier it is to hide the costs of switching between power states. Software mechanisms can help consolidate resource use by moving data and computation across resources. In that case, the energy savings from consolidation should exceed the energy overhead of such movements.

For consolidating resource use across time, previous work on energy-efficient prefetching and caching for mobile computing proposed modifications to the OS to encourage burstiness and increase the length of idle periods [PS04]. A database storage manager could also incorporate similar techniques, especially since certain table scans have highly predictable access patterns. At a higher level, when considering entire systems or collections of resources, we expect to see workload management policies that encourage identifiable periods of low and high activity – perhaps batching requests at the cost of increased latency.

To date, numerous techniques have successfully targeted idle periods to improve pure performance of data management systems. In this case, the goal was to move computation and data to increase overlap among occupied resources. As an example, asynchronous I/O is a mechanism that falls under this category. Energy-efficient techniques that aim to consolidate resource use across time can potentially utilize similar mechanisms.

Further, we could imagine buffer and storage management policies that move data across memory and disks to consolidate space-shared resources. This consolidation would enable powering down unused hardware at the expense of data movement. We could imagine performing these at a coarse granularity, just as load-balancing techniques move database state to improve performance.

## 4.3 Redesign for Max Energy Efficiency

*Approach: Redesign software components to (a) minimize energy use, (b) reduce code bloat, and (c) sacrifice certain properties (or allow underperform in certain metrics) to improve energy efficiency.*

Certain power-oriented tradeoffs in the usage of various resources will not be achievable through existing configuration parameters and thus, we expect certain approaches to energy efficiency to require significant modifications at the software layer. In general, redesigning system components to include faster algorithms and mechanisms improves not only performance but also energy efficiency. However, several algorithms will need to be specifically redesigned for energy use. Consider, for example, the buffer manager: its whole notion and associated replacement policies are based on avoiding as much as possible costly (in terms of latency) accesses to slower storage. With energy savings in mind, the access costs of memory hierarchy levels are going to be different. Moreover, keeping a page in RAM will require energy, proportional to the time the page is cached. New caching and replacement policies will be needed, possibly involving a larger number of more diverse memory hierarchy levels.

Software engineering practices will also need to be revisited and rethought. Multiple layers of abstraction in the code structure along with general-purpose component functionality have led to increases in programmer productivity but also contributed to significant bloat in the code [MS07]. Useful data is typically stored with significant structural overhead, and simple operations involve the execution of a disproportional large set of instructions, creating multiple temporary objects on the way. While from a performance point of view it may not worth addressing code bloat, energy efficiency considerations may reverse that.

In recent years, there has been a rising interest in database-like systems that do not include the full suite of traditional database features. These new systems were designed around new applications with a different set of tradeoffs. Sacrificing certain properties such as consistency, reliability, availability, or even security, can lead to an interesting set of tradeoffs and therefore to opportunities for improving performance. We expect this kind of tradeoffs and reduced-functionality designs to also apply to energy optimizations.

## 5. FUTURE DATABASE DIRECTIONS

Having discussed some general software approaches to reduce energy waste, in this section we point out areas in data management systems that are promising for energy-efficiency optimizations.

## 5.1 Data Placement and Query Processing

**Physical database design**. Decisions on how and where data is stored are expected to have a significant impact on database energy use since initial studies show that more than half the power use is concentrated in the disk subsystem [RSR+07, PN+08]. In addition to re-evaluating all previous tradeoffs of redundant storage and its cost, the new challenge is to take advantage of more choices of physical locations for storing data and incorporate those into the design process. For example, with performance in mind, the physical locations for permanently storing data were restricted to only disks (that are uniformly fast), and, more recently, to faster, albeit more expensive, solid state drives. With energy efficiency in mind, we expect to see more choices: different sets of disk arrays that vary in performance/power characteristics, different types of solid state drives, along with remote storage, accessible over a network. Furthermore, for read-mostly workloads, increasing redundancy may improve energy efficiency. Additional capacity on disks does not carry energy costs if the disk usage remains the same, and, as the example of Section 3.1 hinted, different partitioning schemes may be optimal for different workloads. Lastly, techniques that reduce disk bandwidth requirements, such as column-oriented storage and compression, will need to be re-evaluated for their ability to reduce overall energy use.

**Query optimization and processing algorithms**. The same way query optimization and query processing algorithms are crucial to absolute performance, we expect them to also play a central role in reducing energy waste. Current query processing algorithms are based on fundamental assumptions regarding the size of available memory, the nature and number of accesses they make to both main memory and secondary storage, and their CPU requirements. Optimizing for energy use will first bring changes to the implementation of the algorithms themselves, but most importantly it will change the way the query optimizer estimates costs and chooses a query plan, as we mention in Section 4.1.

## 5.2 Resource Managers

**Query scheduling and memory management**. Due to the complexity of database systems and the wide variety of resources needed by a query over its lifetime (along with the complications that arise when multiplexing the execution of several queries), there has been a limited number of query scheduling policies that work well in practice. The objective in the past has been to improve query response times (either for individual queries or groups of queries) and maximize the utilization of available system resources. For complex queries, scheduling the various operators within a query and deciding on resource allocation for each operator has been crucial for the system's overall performance.

Again, switching objectives, from performance to energy efficiency warrants a thorough re-examination of work done in scheduling and resource management for database workloads. Techniques that enable and encourage work sharing across queries will become increasingly attractive.

**Buffer manager and storage manager**. As we mention is Section 4.3, we expect the buffer and storage management policies to be significantly revised, to reflect energy costs for accessing and storing data, as well as leverage new energy-efficient levels in the memory hierarchy.

**Logging and recovery**. While the actual process of system recovery is not likely to be a candidate for improving energy efficiency (since it happens rarely enough, that performance and correctness will still be the focus), a lot of a system's resources, both in computational power and use of the storage hierarchy, are devoted to logging: previous work showed that about 15% of the code executed in an online transaction processing system is related to logging [HAM+08]. Switching the optimization objectives and incorporating new technology (e.g., flash memories) will affect the logging mechanisms. For example, it may make sense to increase the batching factor (and increase response time) to avoid frequent commits on stable storage, or it might make sense to migrate certain data and transactions to operate directly on stable storage and therefore reduce the amount of logging needed.

## 5.3 New System Architectures

**New implementations**. Traditional database systems are packed with functionality to appease the mid-range of the database market. In the past, this feature-driven approach has increased system complexity (i.e., bloat) and limited performance and scalability. As a result, we are seeing the development of data management software specialized for certain domains, e.g., data warehousing, streaming, transaction processing, application log processing, etc. Similarly, we expect that software complexity may limit the feasibility of energy optimizations. For example, a database vendor may prefer not to implement online data repartitioning because it may affect the usability of numerous external tools, including those distributed by partners. Further, we expect that different energy optimizations are applicable in different domains, e.g., SSDs are better suited for transactional application rather than warehousing. As a result, we expect the trend toward specialization to continue and energy reduction techniques to emerge first from more narrowly focused implementations.

Many of the recent redesigns for improved performance have done more than separate out functionality. Some have also sacrificed other metrics such as consistency for improved performance. Similarly, we expect optimizations that sacrifice availability, reliability, or other "-abilities" to further improve energy use.

**Co-design.** In redesigning data management software, we believe co-design with other disciplines will enable energy efficiency improvements that cannot be attained in isolation. We discuss two important opportunities for co-design.

First, as in the past, data management considerations have influenced the design of hardware (e.g., CPUs and GPGPUs) for improved performance and should do so in the energy space. Our community should influence power-management and hardware architects to develop technologies that address the energy bottlenecks found in data management systems. Otherwise, we will be left optimizing for components that hardly make a difference. Conversely, we will also need to anticipate and adapt our algorithms to the multitude of technologies architects develop to address the larger market such as heterogeneous multicores, different types of solid state technologies (Flash, phase-change RAM, etc.), multi-speed drives, and so on.

A second opportunity is in the coordination of power management performed at the database system level with other independent power management controllers at different levels of the system. For example, consider a hardware controller that changes the voltage and frequency in parallel with the query optimizer which is making decisions based on current runtime power states. If these two do not communicate and coordinate their choices, they may end up working cross purposes [RRT+08]. The software needs to ensure there is an efficient handoff from one controller to another, and ideally, it needs to implement an architecture that helps communicate information across controllers to reach a global optima. To achieve this, database designers will need broad cooperation from people across a variety of disciplines from computer architecture to operating systems to control theory.

**Designing for Total Cost of Ownership.** Although the ratios vary across installations, data center operators recognize management, hardware, and energy costs as the three main costs of the total cost of ownership (TCO). Since energy costs are rising and hardware costs are dropping relatively, we speculate that there will eventually be an opportunity in redesign to sacrifice hardware cost for improved energy efficiency.

For example, in configuring a system for maximum energy efficiency, we may end up with an configuration that does not meet minimum performance criteria. Two potential solutions for increased performance are to either waste energy and increase performance with diminishing returns or pay for more hardware (use more resources in a cluster) and parallelize, keeping the same energy efficiency. Over time, we expect that the latter solution will prevail since the energy costs will make up a larger fraction of TCO. In that respect, we speculate that parallelization and system scalability will continue to be important avenues for maintaining maximum efficiency.

## 6. CONCLUSIONS

Energy efficiency is quickly emerging as a critical research topic across several disciplines. In this paper, our intention was to bring awareness to the database systems community about the opportunities and challenges of energy-aware database computing. We argued that static hardware configurations, energy-proportional hardware, and application-agnostic power management techniques are only part of the answer in controlling and reducing rising energy costs in large data centers. Data management software will ultimately play a significant role in optimizing for energy efficiency. Towards this goal we discussed possible solution approaches as well as areas in database systems ripe for energy improvements. Looking ahead, we urge the database systems community to take up on this challenge and shift focus from performance-oriented research to energy-efficient computing.

## 7. ACKNOWLEDGMENTS

We thank Goetz Graefe for his valuable comments and constructive participation in discussions on energy-efficient software.